\documentclass[pra, aps, twocolumn, preprintnumbers,showpacs, superscriptaddress]{revtex4}
\usepackage{hyperref}
\usepackage{latexsym}
\usepackage{amsmath}

\usepackage{amsfonts}

\def\Complex{\mathbb{C}}
\def\Natural{\mathbb{N}}
\def\id{\mathbb{I}}
\def\01{\{0,1\}}

\newcommand{\ket}[1]{|#1\rangle}
\newcommand{\bra}[1]{\langle#1|}
\newcommand{\outp}[2]{|#1\rangle\langle#2|}
\newcommand{\inp}[2]{\langle#1|#2\rangle}

\newcommand{\N}{\mbox{N}}
\newcommand{\MOLS}{\mbox{MOLS}}

\newcommand{\mub}{b}
\newcommand{\td}{\tau}
\newcommand{\ran}{\rangle}
\newcommand{\lan}{\langle}
\newcommand{\Cn}{\mathbb{C}}
\newcommand{\be}{\begin{eqnarray} \begin{aligned}}
\newcommand{\ee}{\end{aligned} \end{eqnarray} }
\newcommand{\benn}{\begin{eqnarray*} \begin{aligned}}
\newcommand{\eenn}{\end{aligned} \end{eqnarray*} }

\newcommand{\Tr}{\mathop{\mathrm{Tr}}\nolimits}
\newcommand{\cmb}[2]{\left(\begin{array}{@{}c@{}}#1\\#2\end{array}\right)}
\newcommand{\mV}{\mathcal{V}}

\newcommand{\hil}{\mathcal{H}}
\newcommand{\U}{\textrm{U}}

\newtheorem{definition}{Definition}
\newtheorem{theorem}{Theorem}
\newtheorem{lemma}{Lemma}

\newtheorem{corollary}{Corollary}

\newtheorem{claim}{Claim}

\newenvironment{proof}
{\noindent {\bf Proof. }}
{{\hfill $\Box$}\\
 \smallskip}

\newcommand{\mB}{\mathcal{B}}

\newcommand{\mI}{\mathcal{I}}
\newcommand{\mS}{\mathcal{S}}
\newcommand{\ens}{\mathcal{E}}

\begin{document}
\title{Entropic uncertainty relations and locking: tight bounds for mutually unbiased bases}
\author{Manuel A. \surname{Ballester}}
\email[]{Manuel.Ballester@cwi.nl}
\affiliation{Centrum voor Wiskunde en Informatica, Kruislaan 413, 1098 SJ
Amsterdam, The Netherlands}
\author{Stephanie \surname{Wehner}}
\email[]{wehner@cwi.nl}
\affiliation{Centrum voor Wiskunde en Informatica, Kruislaan 413, 1098 SJ
Amsterdam, The Netherlands}

\begin{abstract}
We prove tight entropic uncertainty relations for a large number of mutually unbiased measurements. 
In particular, we show that a bound derived from the result by 
Maassen and Uffink~\cite{maassen:entropy} for 2 such measurements can in fact be tight for up 
to $\sqrt{d}$ measurements in mutually unbiased bases.
We then show that using more mutually unbiased bases does not always lead to
a better locking effect. We prove that the optimal bound
for the accessible information using up to $\sqrt{d}$ specific mutually unbiased
bases is $\log d/2$, which is the same as can be achieved by using only two bases.
Our result indicates that merely using mutually unbiased bases is not
sufficient to achieve a strong locking effect, and we need to look for
additional properties.
\end{abstract}

\maketitle

We investigate two related notions that are of importance in many
quantum cryptographic tasks: entropic uncertainty relations and locking classical information in quantum states.
\smallskip\newline
\noindent \textbf{Entropic uncertainty relations} are an 
alternative way to state Heisenberg's uncertainty principle.
They are frequently a more useful characterization, because the ``uncertainty'' is lower
bounded by a quantity that does not depend on the state to be measured~\cite{deutsch:uncertainty,kraus:entropy}.
Recently, entropic uncertainty relations have gained importance in the context of quantum cryptography
in the bounded storage model, where proving the security of such protocols ultimately
reduces to bounding such relations~\cite{serge:newthing}. Proving new entropic uncertainty relations
could thus give rise to new protocols. 
Such relations are known for two~\cite{maassen:entropy},
or $d+1$~\cite{sanchez:entropy2,sanchez:entropy3} mutually unbiased measurements (see Section~\ref{prelim} for a
definition).
Very little, however,
is known for any other number of measurements~\cite{azarchs:entropy}.

Here, we prove tight entropic uncertainty relations for measurements in a large number of 
mutually unbiased bases (MUBs) in square dimensions. In particular, we consider any MUBs derived
from mutually orthogonal Latin squares~\cite{wocjan:mub}, and \emph{any} set of MUBs 
obtained from the set of unitaries of the form $\{U \otimes U^*\}$, where $\{U\}$ gives 
set of MUBs in dimension $s$ when applied to the basis elements of the computational basis. 
For any $s$, there are at most $s+1$ such MUBs in a Hilbert space of dimension $d=s^2$.
Let $\mathbb{B}$ be the set of MUBs coming from one of these two constructions. 
We prove that for any subset $\mathbb{T} \subseteq \mathbb{B}$ of these bases we
have 
$$ 
\min_{\ket{\phi}} \sum_{\mB \in \mathbb{T}} H(\mB,\ket{\phi}) = \frac{|\mathbb{T}|}{2} \log d,
$$
where $H(\mB,\ket{\phi}) =  - \sum_{i = 1}^{d}
|\inp{\phi}{b_i}|^2 \log |\inp{\phi}{b_i}|^2$ is the Shannon entropy~\cite{shannon:info} arising from 
measuring the state $\ket{\phi}$ in the basis $\mB = \{|b_1\ran,\ldots,|b_d\ran\}$.

Our result furthermore shows that one needs to be careful to think of ``maximally incompatible'' measurements
as being necessarily mutually unbiased. When we take entropic uncertainty relations as our measure of
``incompatibility'', mutually unbiased measurements are in fact not always the most incompatible when
considering more than two observables. In particular, it has been shown \cite{winter:randomizing} that if we choose approximately
$(\log d)^4$ bases uniformly at random, then 
$\min_{\ket{\phi}} (1/|\mathbb{T}|) \sum_{\mB \in \mathbb{T}} H(\mB,\ket{\phi}) \geq \log d -3$. This means that
there exist $(\log d)^4$ bases for which this sum of entropies is very large, i.e., measurements in such bases
are very incompatible. However, we showed that when $d$ is large, there exist $\sqrt{d}$, 
mutually unbiased bases which are much less incompatible according to this measure.
When considering entropic uncertainty relations as a measure of ``incompatibility'', we must therefore
look for different properties for the bases to define incompatible measurements.

Finally, we give an alternative proof that if $\mathbb{B}$ is a 
set of $d+1$ MUBs we have 
$\sum_{\mB \in \mathbb{B}} H(\mB,\ket{\phi}) \geq (d+1) \log((d+1)/2)$~\cite{sanchez:entropy2}.
Our proof is based on the fact that such a set forms a 2-design,
which may offer new insights. 
\smallskip\newline
\noindent\textbf{Locking} classical correlations in quantum states is an exciting feature
of quantum information~\cite{terhal:locking}, intricately related to entropic uncertainty relations. 
Consider a two-party 
protocol with one or more rounds of communication. Intuitively, 
one would expect that in each round the amount of correlation between
the two parties cannot increase by much more than the amount of data transmitted.
For example, transmitting  $2\ell$ classical bits or $\ell$ qubits (and using superdense coding) 
should not increase the amount of correlation by more than $2\ell$ bits,
no matter what the initial state of the two party system was. This intuition
is accurate when we take the classical mutual information $\mI_c$ as
our correlation measure, and require all communication to be classical.
However, when quantum communication is possible at some point
during the protocol, everything changes: there exist two-party mixed quantum 
states, such that transmitting just a 
single extra bit of classical communication can result in an arbitrarily large 
increase in $\mI_c$~\cite{terhal:locking}. 
The magnitude of this increase thereby
only depends on the dimension of the initial mixed state. Since then, similar locking
effects have been observed also for other correlation
measures~\cite{m:locking,karol:locking}. Such effects play a role in very different
scenarios: they have been used to explain physical phenomena related to black holes~\cite{jsmo:locking}, but
they are also important in crypto\-gra\-phic applications such as quantum key distribution~\cite{rk:locking}
and quantum bit string commitment~\cite{qsc,qscPRL}. We are thus interested in
determining how exactly we can obtain locking effects, and how dramatic they can be.

The correlation measure considered here, is the classical mutual
information of a bipartite quantum state $\rho_{AB}$, which is the maximum classical 
mutual information that can be obtained by local measurements $M_A \otimes M_B$
on the state $\rho_{AB}$~\cite{terhal:minfo}:
\begin{equation}\label{mutualInfo}
\mI_c(\rho_{AB}) = \max_{M_A \otimes M_B} \mI(A:B).
\end{equation}
The classical mutual information is defined as 
$\mI(A:B) = H(P_A) + H(P_B) - H(P_{AB})$
where $H$ is the Shannon entropy. $P_A$, $P_B$, and $P_{AB}$ are
the probability distributions corresponding to the individual and joint outcomes
of measuring the state $\rho_{AB}$ with $M_A \otimes M_B$. 
The mutual information between $A$ and $B$ is a measure of the information
that $B$ contains about $A$.
This measure of correlation is of particular relevance for quantum bit string commitments~\cite{qsc,qscPRL}.
Furthermore, the first locking effect was observed for this quantity 
in the following protocol 
between two parties: Alice (A) and Bob (B). 
Let $\mathbb{B} = \{\mB_1,\ldots,\mB_m\}$ with $\mB_t = \{\ket{b^t_1},\ldots,\ket{b^t_d}\}$
be a set of $m$ MUBs in $\Complex^d$.
Alice picks an
element $k \in \{1,\ldots,d\}$ and a basis $\mB_t \in \mathbb{B}$ uniformly at random. She then
sends $\ket{b^t_k}$ to Bob, while keeping $t$ secret. 
Such a protocol gives 
rise to the joint state
$$
\rho_{AB} = \frac{1}{m d} 
\sum_{k=1}^{d} \sum_{t=1}^{m} (\outp{k}{k} \otimes \outp{t}{t})_A
\otimes (\outp{b_k^t}{b^t_k})_B.
$$
Clearly, if Alice told her basis choice $t$ to Bob, 
he could measure in the right basis and obtain the correct $k$. 
Alice and Bob would then share $\log d + \log m$ bits of correlation,
which is also their mutual information $\mI_c(\sigma_{AB})$,
where $\sigma_{AB}$ is the state obtained from $\rho_{AB}$ after the announcement of $t$. 
But, how large is $\mI_c(\rho_{AB})$,  when Alice does \emph{not} announce $t$ to Bob? 
It was shown~\cite{terhal:locking} that in dimension $d=2^n$, using the two MUBs given by 
the unitaries $\id^{\otimes n}$ and $H^{\otimes n}$ applied to the computational basis, where $H$ is the Hadamard matrix,
we have $\mI_c(\rho_{AB}) = (1/2) \log d$. This means 
that the single bit of basis information Alice transmits to Bob,
``unlocks'' $(1/2) \log d$ bits: \emph{without} this bit, the mutual information
is $(1/2) \log d$, but \emph{with} this bit it is $\log d + 1$.
It is also known that if Alice
and Bob randomly choose a large set of unitaries from the Haar
measure to construct $\mathbb{B}$, then $\mI_c$
can be brought down to a small constant~\cite{winter:randomizing}.
However, no explicit constructions with more than two bases are known 
that give good locking effects. Based on numerical studies for
spaces of prime dimension $3 \leq d \leq 30$, one might hope that adding
a third MUB would strengthen the locking effect 
and give $\mI_c(\rho_{AB}) \approx (1/3) \log d$~\cite{terhal:locking}.

Here, however, we show that this intuition fails us. We 
prove that for three MUBs given by
$\id^{\otimes n}$, $H^{\otimes n}$, and $K^{\otimes n}$ where $K = (\id + i\sigma_x)/\sqrt{2}$ and 
dimension $d=2^n$ for some even integer $n$, we have 
\begin{equation}\label{minf}
\mI_c(\rho_{AB}) = (1/2) \log d,
\end{equation}
the same locking effect as with two MUBs. 
We also show that for any subset of the MUBs based on Latin squares and the MUBs in square
dimensions based on generalized Pauli matrices~\cite{boykin:mub}, we again obtain (\ref{minf}), i.e., using
two or all $\sqrt{d}$ of them makes no difference at all.
Finally, we show that for any set of MUBs $\mathbb{B}$ based on generalized Pauli matrices in \emph{any} dimension,
$\mI_c(\rho_{AB}) = \log d - \min_{\ket{\phi}}(1/|\mathbb{B}|)\sum_{\mB \in \mathbb{B}}H(\mB,\ket{\phi})$, i.e., it 
is enough to determine a bound on the entropic uncertainty relation to determine the 
strength of the locking effect.
Although bounds for general MUBs still elude us, our results show
that merely choosing the bases to be mutually unbiased is not sufficient and we must look
elsewhere to find bases which provide good locking. 

\section{Preliminaries}\label{prelim}

Throughout this paper, we use the shorthand notation $[d] = \{1,\ldots,d\}$. 
We write 
\benn
H(\mB_t,\ket{\phi}) =  - \sum_{i = 1}^{d}|\inp{\phi}{b_k^t}|^2 \log |\inp{\phi}{b_k^t}|^2,
\eenn 
for the Shannon entropy~\cite{shannon:info} 
arising from measuring the pure state $\ket{\phi}$ in basis $\mB_t = \{\ket{b_1^t},\ldots,\ket{b_d^t}\}$.
In general, we will use $\ket{b_k^t}$ with $k \in [d]$ to denote the $k$-th element of a basis $\mB_t$ 
indexed by $t$. We also briefly refer to the R\'enyi entropy of order 2 (collision entropy) 
of measuring $\ket{\phi}$ in basis $\mB_t$ given by $H_2(\mB_t,\ket{\phi}) = - \log \sum_{i=1}^d
|\inp{\phi}{b_k^t}|^4$~\cite{cachin:renyi}.

\subsection{Mutually unbiased bases}

We also need the notion of mutually unbiased bases (MUBs),
which were initially introduced in the context
of state estimation~\cite{wootters:mub}, but appear in many other problems in quantum information.
The following definition closely follows the one given in~\cite{boykin:mub}.
\begin{definition}[MUBs] \label{def-mub}
Let $\mathcal{B}_1 = \{|\mub^1_1\ran,\ldots,|\mub^1_{d}\ran\}$ and $\mB_2 =
\{|\mub^2_1\ran,\ldots,|\mub^2_{d}\ran\}$ be two orthonormal bases in
$\Complex^d$. They are said to be
\emph{mutually unbiased}  if
$|\lan\mub^1_k |\mub^2_l\ran| = 1/\sqrt{d}$, for every $k,l \in[d]$. A set $\{\mathcal{B}_1,\ldots,\mathcal{B}_m\}$ of
orthonormal bases in $\Cn^d$ is called a \emph{set of mutually
unbiased bases} if each pair of bases is mutually unbiased.
\end{definition}
We use $N(d)$ to denote the maximal number of MUBs in dimension $d$. 
In any dimension $d$, we have that
$\N(d) \leq d+1$~\cite{boykin:mub}. If $d = p^k$ is a prime power, we have
that $\N(d) = d+1$ and explicit constructions are
known~\cite{boykin:mub,wootters:mub}. If $d = s^2$ is a square, 
$\N(d) \geq \MOLS(s)$ where $\MOLS(s)$ denotes the number of mutually orthogonal
$s \times s$ Latin squares~\cite{wocjan:mub}. In general, we have
$\N(n m) \geq \min\{\N(n),\N(m)\}$ for all $n,m \in \Natural$~\cite{zauner:diss,klappenecker:mubs}. 
It is also known that in any dimension, there exists an explicit construction for 3 MUBs~\cite{grassl:mub}.
Unfortunately, not very much is known for other dimensions. For example,
it is still an open problem whether there exists a set of $7$ MUBs in dimension $d=6$.
We say that a unitary $U_t$ transforms the computational basis into the $t$-th MUB 
$\mB_t = \{\ket{b^t_1},\ldots,\ket{b^t_d}\}$ 
if for all $k \in [d]$ we have $\ket{b^t_k} = U_t\ket{k}$.
Here, we are particularly concerned with two specific constructions of mutually unbiased bases.

\subsubsection{Latin squares}

First of all, we consider MUBs based on mutually orthogonal Latin squares~\cite{wocjan:mub}. 
Informally, an $s \times s$ Latin square over the symbol set $[s] = \{1,\ldots,s\}$ is an arrangement
of elements of $[s]$ into an $s \times s$ square such that in each row and each column every element
occurs exactly once. Let $L_{ij}$ denote the entry in a Latin square in row $i$ and column $j$. 
Two Latin squares $L$ and $L'$ are called mutually orthogonal if and only if
$\{(L_{i,j},L'_{i,j})|i,j \in [s]\} = \{(u,v)|u,v \in [s]\}$. 
From any $s\times s$ Latin square we can obtain a basis for $\Cn^{s}\otimes \Cn^{s}$.  
First, we construct $s$ of the basis vectors from the entries of
the Latin square itself. Let $\ket{v_{1,\ell}} = (1/\sqrt{s}) \sum_{i,j\in [s]} E^L_{i,j}(\ell) \ket{i,j}$
where $E^L$ is a predicate such that $E^L_{i,j}(\ell) = 1$ if and only if $L_{i,j} = \ell$.
Note that for each $\ell$ we have exactly $s$ pairs $i,j$ such that $E_{i,j}(\ell) = 1$, because
each element of $[s]$ occurs exactly $s$ times in the Latin square.
Secondly, from each such vector we obtain $s-1$ additional vectors by adding successive rows
of an $s \times s$ (complex) Hadamard matrix $H = (h_{ij})$ as coefficients to obtain the remaining 
$\ket{v_{t,j}}$ for $t \in [s]$, where $h_{ij} = \omega^{ij}$ with $i,j \in \{0,\ldots,s-1\}$ and
$\omega = e^{2 \pi i/s}$.
Two additional MUBs can then be obtained in the same way from the two non-Latin squares where
each element occurs for an entire row or column respectively. From each mutually orthogonal latin square
and these two extra squares which also satisfy the above orthogonality condition, we obtain one basis.
This construction therefore gives $\MOLS(s) + 2$ many MUBs. It is known that if $s = p^k$ is a 
prime power itself, we obtain
$p^k+1\approx \sqrt{d}$ MUBs from this construction. Note, however, that there do exist many more
MUBs in prime power dimensions, namely $d+1$. If $s$ is not a prime power, it is merely known
that $\MOLS(s) \geq s^{1/14.8}$~\cite{wocjan:mub}.

As an example, consider the following $3 \times 3$ Latin square and
the $3 \times 3$ Hadmard matrix\\
\begin{center}
\begin{tabular}{lr}
\begin{tabular}{|c|c|c|}
\hline
1 & 2 & 3\\\hline
2 & 3 & 1\\\hline
3 & 1 & 2 \\\hline
\end{tabular}, 
&
$
H = \left(\begin{array}{ccc}
    1 &1& 1\\
    1 &\omega &\omega^2\\
    1 &\omega^2& \omega
\end{array}\right)$,
\end{tabular}
\end{center}
where $\omega = e^{2 \pi i/3}$. 
First, we obtain vectors
\begin{eqnarray*}
\ket{v_{1,1}} &=& (\ket{1,1} + \ket{2,3} + \ket{3,2})/\sqrt{3}\\
\ket{v_{1,2}} &=& (\ket{1,2} + \ket{2,1} + \ket{3,3})/\sqrt{3}\\
\ket{v_{1,3}} &=& (\ket{1,3} + \ket{2,2} + \ket{3,1})/\sqrt{3}.
\end{eqnarray*}
With the help of $H$ we obtain 3 
additional vectors from the ones above. From the vector $\ket{v_{1,1}}$,
for example, we obtain 
\begin{eqnarray*}
\ket{v_{1,1}} &=& (\ket{1,1} + \ket{2,3} + \ket{3,2})/\sqrt{3}\\
\ket{v_{2,1}} &=& (\ket{1,1} + \omega \ket{2,3} + \omega^2 \ket{3,2})/\sqrt{3}\\
\ket{v_{3,1}} &=& (\ket{1,1} + \omega^2 \ket{2,3} + \omega \ket{3,2})/\sqrt{3}.
\end{eqnarray*}
This gives us basis $\mathcal{B} = \{\ket{v_{t,\ell}}|t,\ell \in [s]\}$ for $s = 3$.
The construction of another basis follows in exactly the same way from a mutually orthogonal
Latin square. The fact that two such squares $L$ and $L'$ are mutually orthogonal ensures
that the resulting bases will be mutually unbiased. Indeed, suppose we are given another such basis, 
$\mathcal{B'} = \{\ket{u_{t,\ell}}|t,\ell \in [s]\}$ belonging to $L'$. We then have for any $\ell,\ell' \in [s]$ that
$|\inp{u_{1,\ell'}}{v_{1,\ell}}|^2 = 
|(1/s) \sum_{i,j\in [s]} E^{L'}_{i,j}(\ell') E^L_{i,j}(\ell)|^2 = 1/s^2$, as there exists excactly only
one pair $\ell,\ell' \in [s]$ such that $E^{L'}_{i,j}(\ell') E^L_{i,j}(\ell) = 1$. Clearly, the same argument
holds for the additional vectors derived from the Hadamard matrix.
 
\subsubsection{Generalized Pauli matrices}

The second construction we consider is based on the generalized Pauli matrices 
$X_d$ and $Z_d$~\cite{boykin:mub}, defined by their actions on the 
computational basis $C = \{\ket{1},\ldots,\ket{d}\}$ as follows: 
$$
X_d\ket{k} = \ket{k+1}, \and Z_d\ket{k} = \omega^k\ket{k},~\forall \ket{k} \in C,
$$
where $\omega = e^{2 \pi i/d}$.  We say that $\left(X_{d}\right)^{a_1} \left(Z_{d}\right)^{b_1} 
\otimes \cdots \otimes \left(X_{d}\right)^{a_N} \left(Z_{d}\right)^{b_N}$ for 
$a_k,b_k \in \{0,\ldots,d-1\}$ and $k \in [N]$ is a \emph{string of Pauli Matrices}.

If $d$ is a prime, it is known that the $d+1$ MUBs constructed first by
Wootters and Fields~\cite{wootters:mub} can also be obtained as the eigenvectors of the 
matrices $Z_d,X_d,X_dZ_d,X_dZ_d^2,\ldots,X_dZ_d^{d-1}$~\cite{boykin:mub}. If $d = p^k$ is a prime power,
consider all $d^2-1$ possible strings of Pauli matrices excluding the identity and group them
into sets $C_1,\ldots,C_{d+1}$ such that $|C_i| = d - 1$ and $C_i \cup C_j = \{\id\}$ for $i \neq j$ and all
elements of $C_i$ commute. Let $B_i$ be the common eigenbasis of all elements of $C_i$. Then
$B_1,\ldots,B_{d+1}$ are MUBs~\cite{boykin:mub}. A similar result for $d = 2^k$ has also been shown 
in~\cite{lawrence:mub}.
A special case of this construction are the three mutually unbiased bases in dimension $d=2^k$ given by 
the unitaries $\id^{\otimes k}$,$H^{\otimes k}$ and $K^{\otimes k}$ with $K = (\id + i\sigma_x)/\sqrt{2}$ applied to the computational
basis. 

\subsection{2-designs}

For the purposes of the present work, \emph{spherical $t$-designs} (see for example Ref.\ \cite{Renesetal04a}) can be defined as follows.

\begin{definition}[$t$-design]
Let $\{|\td_1\ran,\ldots,|\td_{m}\ran\}$ be a set of state vectors in $\Cn^d$, they are said to form a $t$-design if 
\benn
\frac{1}{m}\sum_{i=1}^m [|\td_i\ran \lan \td_i|]^{\otimes t} = \frac{\Pi_+^{(t,d)}}{\Tr \Pi_+^{(t,d)}},
\eenn
where $\Pi_+(t,d)$ is a projector onto the completely  symmetric subspace of ${\Cn^d}^{\otimes t}$ and 
\benn
\Tr \Pi_+^{(t,d)}=\cmb{d+t-1}{d-1}=\frac{(d+t-1)!}{(d-1)!~t!},
\eenn
is its dimension.
\end{definition}
Any set $\mathbb{B}$ of $d+1$ MUBs forms a \emph{spherical $2$-design}  \cite{KlappeneckerRotteler05a,Renesetal04a}, 
i.e., we have for $\mathbb{B} = \{\mB_1,\ldots,\mB_{d+1}\}$ with $\mB_t = \{\ket{b^t_1},\ldots,\ket{b^t_d}\}$ that
\benn
\frac{1}{d(d+1)}\sum_{t=1}^{d+1}\sum_{k=1}^{d} [|\mub^t_k\ran \lan \mub^t_k|]^{\otimes 2} &= 2\frac{\Pi_+^{(2,d)}}{d(d+1)}.
\eenn

\section{Uncertainty relations}

We now prove tight entropic uncertainty for measurements in MUBs in square dimensions. 
The main result of~\cite{maassen:entropy}, which will be very useful for us, is stated next.
\begin{theorem}[Maassen and Uffink]
Let $\mB_1$ and $\mB_2$ be two orthonormal basis in a Hilbert space of dimension $d$. Then
for all pure states $\ket{\psi}$
\be \label{eq:maasenuffinkbound}
\frac{1}{2}\left[ H(\mB_1,\ket{\psi})+H(\mB_2,\ket{\psi})\right]\geq -\log c(\mB_1,\mB_2),
\ee
where $c(\mB_1,\mB_2)=\max \left \{|\lan b_1|b_2\ran|:|b_1\ran \in \mB_1,|b_2\ran \in \mB_2\right \}$.
\end{theorem}
The case when $\mB_1$ and $\mB_2$ are MUBs is of special interest for us. More generally, when one has a set of MUBs a trivial application of (\ref{eq:maasenuffinkbound}) leads to the following corollary also noted in~\cite{azarchs:entropy}.
\begin{corollary}\label{MUderived}
Let $\mathbb{B}=\{\mB_1,\ldots,\mB_m\}$, be a set of MUBs in a Hilbert space of dimension $d$. Then
\be \label{eq:manymubsbound}
\frac{1}{m} \sum_{t=1}^m H(\mB_t,|\psi\ran)\geq \frac{\log d}{2}.
\ee
\end{corollary}
\begin{proof}
Using (\ref{eq:maasenuffinkbound}), one gets that for any pair of MUBs $\mB_t$ and $\mB_{t'}$ with $t\neq t'$
\be\label{eq:maassenuffinkij}
\frac{1}{2}\left[ H(\mB_t,\psi)+H(\mB_{t'},\psi)\right]\geq \frac{\log d}{2}.
\ee
Adding up the resulting equation for all pairs $t\neq t'$ we get the desired result (\ref{eq:manymubsbound}).
\end{proof}
Here, we now show that this bound can in fact be tight for a large set of MUBs. 

\subsection{MUBs in square dimensions}
Corollary \ref{MUderived}, gives  a lower bound on the average of the entropies of a set of MUBs. The obvious question is whether that bound is tight. We show that the bound is indeed tight when we consider product MUBs in a Hilbert space of square dimension.
\begin{theorem}\label{squareThm}
Let $\mathbb{B}=\{\mB_1,\ldots,\mB_m\}$ with $m\geq 2$ be a set of MUBs in a Hilbert space $\hil$ of dimension $s$. Let $U_t$ be the unitary operator that transforms the computational basis to $\mB_t$. 
Then $\mathbb{V}=\{\mV_1,\ldots,\mV_m\}$, where 
\benn
\mV_t=\left \{U_t|k\ran \otimes U_t^* |l\ran: k,l\in[s] \right \},
\eenn
is a set of MUBs in $\hil \otimes \hil$, and it holds that
\be\label{eq:squaremubsbound}
\min_{\ket{\psi}} \frac{1}{m} \sum_{t=1}^m H(\mV_t,|\psi\ran)= \frac{\log d}{2},
\ee
where $d=\dim(\hil \otimes \hil)=s^2$. 
\end{theorem}
\begin{proof}
It is easy to check that $\mathbb{V}$ is indeed a set of MUBs. Our proof works by constructing a state $\ket{\psi}$
that achieves the bound in Corollary~\ref{MUderived}.
It is easy to see that the maximally entangled state
\benn
|\psi\ran = \frac{1}{\sqrt{s}}\sum_{k=1}^{s}|kk\ran,
\eenn
satisfies $U\otimes U^*|\psi\ran=|\psi\ran$ for any $U\in \U(d)$. Indeed,
\benn
\lan \psi |U\otimes U^*|\psi\ran&=\frac{1}{s}\sum_{k,l=1}^{s} \lan k |U|l\ran\lan k |U^*|l\ran\\
						&=\frac{1}{s}\sum_{k,l=1}^{s} \lan k |U|l\ran\lan l |U^\dagger|k\ran\\
						&=\frac{1}{s}\Tr UU^\dagger=1. 
\eenn
Therefore, for any $t\in[m]$ we have that
\benn
H(\mV_t,|\psi\ran)	&=-\sum_{kl}|\lan kl|U_t\otimes U_t^*|\psi\ran|^2\log|\lan kl|U_t\otimes U_t^*|\psi\ran|^2\\
				&=-\sum_{kl}|\lan kl|\psi\ran|^2\log|\lan kl|\psi\ran|^2\\
				&=\log s=\frac{\log d}{2}.
\eenn
Taking the average of the previous equation we get the desired result.
\end{proof}

\subsection{MUBs based on Latin Squares}

We now consider mutually unbiased bases based on Latin squares~\cite{wocjan:mub} 
as described in Section~\ref{prelim}. Our proof again follows by providing a state
that achieves the bound in Corollary~\ref{MUderived}, which turns out to have
a very simple form. 
\begin{lemma}\label{LSentropy}
Let $\mathbb{B}=\{\mB_1,\ldots,\mB_m\}$ with $m \geq 2$ be any set of MUBs in a Hilbert space of dimension $d=s^2$ constructed
on the basis of Latin squares. Then 
$$
\min_{\ket{\psi}} \frac{1}{m} \sum_{\mB\in\mathbb{B}} H(\mB,\ket{\psi}) = \frac{\log d}{2}.
$$
\end{lemma}
\begin{proof}
Consider the state $\ket{\psi} = \ket{1,1}$ and fix a basis 
$\mB_t = \{\ket{v^t_{i,j}}|i,j \in [s]\} \in \mathbb{B}$ 
coming from a Latin square. 
It is easy 
to see that there exists exactly one $j \in [s]$ such that $\inp{v^t_{1,j}}{1,1} = 1/\sqrt{s}$. Namely this 
will be the $j \in [s]$ at position $(1,1)$ in the Latin square. Fix this $j$. For any other 
$\ell \in [s], \ell \neq j$, we have $\inp{v^t_{1,\ell}}{1,1} = 0$. But this means that 
there exist exactly $s$ vectors in $\mB$ such that $|\inp{v^t_{i,j}}{1,1}|^2 = 1/s$, namely 
exactly the $s$ vectors derived 
from $\ket{v^t_{1,j}}$ via the Hadamard matrix. The same argument holds for any such basis $\mB \in \mathbb{T}$.
We get
\begin{eqnarray*}
\sum_{\mB \in \mathbb{B}} H(\mB,\ket{1,1}) &=& \sum_{\mB \in \mathbb{B}} 
\sum_{i,j \in [s]} |\inp{v^t_{i,j}}{1,1}|^2 \log |\inp{v^t_{i,j}}{1,1}|^2\\
&=& |\mathbb{T}| s \frac{1}{s} \log \frac{1}{s}\\
&=& |\mathbb{T}| \frac{\log d}{2}.
\end{eqnarray*}
The result then follows directly from Corollary~\ref{MUderived}.
\end{proof}

\subsection{Using a full set of MUBs} 

We now provide an alternative proof of an entropic uncertainty relation 
for a full set of mutually unbiased bases. This has previously been proved
in~\cite{sanchez:entropy2}. Nevertheless, because our proof is so 
simple using existing results about 2-designs we include it here for completeness, 
in the hope that if may offer additional insight.
\begin{lemma}\label{FullentropyCollision}
Let $\mathbb{B}$ be a set of $d+1$ MUBs in a Hilbert space of dimension $d$.
Then 
$$
\frac{1}{d+1}\sum_{\mB\in\mathbb{B}}H_2(\mB,\ket{\psi}) \geq \log\left(\frac{d+1}{2}\right).
$$
\end{lemma}
\begin{proof}
Let $\mB_t = \{\ket{b^t_1},\ldots,\ket{b^t_d}\}$ and $\mathbb{B} = \{\mB_1,\ldots,\mB_{d+1}\}$.
We can then write
\begin{eqnarray*}
\frac{1}{d+1}\sum_{\mB\in\mathbb{B}}H_2(\mB,\ket{\psi}) &=&
- \frac{1}{d+1}\sum_{t=1}^{d+1} \log \sum_{k=1}^d |\inp{b^t_k}{\psi}|^4\\
&\geq& \log\left(\frac{1}{d+1}\sum_{t=1}^{d+1}\sum_{k=1}^d |\inp{b^t_k}{\psi}|^4\right)\\
&=&\log\left(\frac{d+1}{2}\right),
\end{eqnarray*}
where the first inequality follows from the concavity of the $\log$,
and the final inequality follows directly from the fact that a full set of MUBs forms a 2-design
and~\cite[Theorem 1]{KlappeneckerRotteler05a}.
\end{proof}

We then obtain the original result by Sanchez-Ruiz~\cite{sanchez:entropy2} by noting that
$H(\cdot) \geq H_2(\cdot)$. 
\begin{corollary}\label{Fullentropy}
Let $\mathbb{B}$ be a set of $d+1$ MUBs in a Hilbert space of dimension $d$.
Then 
$$
\frac{1}{d+1}\sum_{\mB\in\mathbb{B}}H(\mB,\ket{\psi}) \geq \log\left(\frac{d+1}{2}\right).
$$
\end{corollary}

\section{Locking}

We now turn our attention to locking. We first explain the connection between locking
and entropic uncertainty relations. In particular, we show that for MUBs based on generalized
Pauli matrices, we only need to look at such uncertainty relations to determine the exact strength 
of the locking effect. We then consider how good MUBs based on Latin squares are for locking.

In order to determine how large the locking effect is for some set
of mutually unbiased bases $\mathbb{B}$,
and the state
\begin{equation}\label{rhoAB}
\rho_{AB} =
\sum_{t=1}^{|\mathbb{B}|}
\sum_{k=1}^{d} p_{t,k}
(\outp{k}{k} \otimes \outp{t}{t})_A
\otimes (\outp{b^t_k}{b^t_k})_B,
\end{equation}
we must find an optimal bound for $\mI_c(\rho_{AB})$.
Here, $\{p_{t,k}\}$ is a probability distribution over $\mathbb{B} \times [d]$.
That is, we must find a POVM
$M_A \otimes M_B$ that maximizes Eq.\ (\ref{mutualInfo}).
It has been shown in~\cite{terhal:locking} that we can restrict ourselves to to taking
$M_A$ to be the local measurement determined by the projectors $\{\outp{k}{k} \otimes \outp{t}{t}\}$.
It is also known that we can limit ourselves to take the measurement $M_B$ consisting of rank one
elements $\{\alpha_i \outp{\Phi_i}{\Phi_i}\}$ only~\cite{davies:access}, where $\alpha_i \geq 0$ and $\ket{\Phi_i}$
is normalized.
Maximizing over $M_B$ then corresponds to maximizing Bob's accessible information~\cite[Eq.\ (9.75)]{peres:book} for
the ensemble $\ens = \{p_{k,t},\outp{b^t_k}{b^t_k}\}$
\begin{eqnarray}
\begin{aligned}\label{accessible}
&&\mI_{acc}(\ens)= \max_M \left(- \sum_{k,t} p_{k,t} \log p_{k,t} + \right.\\
&&\left.\sum_{i} \sum_{k,t} p_{k,t} \alpha_i \bra{\Phi_i}\rho_{k,t}\ket{\Phi_i}
\log \frac{p_{k,t} \bra{\Phi_i}\rho_{k,t}\ket{\Phi_i}}{\bra{\Phi_i}\mu\ket{\Phi_i}} \right),
\end{aligned}
\end{eqnarray}
where $\mu = \sum_{k,t} p_{k,t} \rho_{k,t}$ and $\rho_{k,t} = \outp{b^t_k}{b^t_k}$. 
Therefore, we have $\mI_c(\rho_{AB}) = \mI_{acc}(\ens)$.
We are now ready to prove our locking results.

\subsection{An example}

We first consider a very simple example with only three MUBs that provides the intuition behind
the remainder of our paper. The three MUBs we consider now are generated by the unitaries
$\id$, $H$ and $K = (\id + i\sigma_x)/\sqrt{2}$ when applied to the computational basis.
For this small example, we also investigate the role of the prior over the bases and the encoded
basis elements. It turns out that this does not affect the strength of the locking effect positively. Actually, it is possible
to show the same for encodings in many other bases. However,we 
do not consider this case in full generality as to not obscure our main line of argument.

\begin{lemma}\label{3mubLocking}
Let $U_0=\id^{\otimes n}$,$U_1 = H^{\otimes n}$, and $U_2 = K^{\otimes n}$, 
where $k \in \{0,1\}^n$ and $n$ is an even integer.
Let $\{p_t\}$ with $t \in [3]$ be a probability distribution over
the set $\mS = \{U_1,U_2,U_3\}$. Suppose that $p_1,p_2,p_3 \leq 1/2$ and let
$p_{t,k} = p_t (1/d)$.
Consider the ensemble $\ens = \{ p_t \frac{1}{d},U_t \outp{k}{k}U_t^\dagger\}$, then
$$
\mI_{acc}(\ens) = \frac{n}{2}.
$$ 
If, on the other hand, there exists a $t \in [3]$ such that
$p_t > 1/2$, then $\mI_{acc}(\ens) > n/2$.
\end{lemma}
\begin{proof}
We first give an explicit measurement strategy and then prove a matching upper bound
on $\mI_{acc}$.
Consider the Bell basis vectors $\ket{\Gamma_{00}} = (\ket{00} + \ket{11})/\sqrt{2}$, 
$\ket{\Gamma_{01}} = (\ket{00} - \ket{11})/\sqrt{2}$, $\ket{\Gamma_{10}} = (\ket{01} + \ket{10})/\sqrt{2}$,
and $\ket{\Gamma_{11}} = (\ket{01} - \ket{10})/\sqrt{2}$. Note that we can write
for the computational basis
\begin{eqnarray*}
\ket{00} &=& \frac{1}{\sqrt{2}}(\ket{\Gamma_{00}} + \ket{\Gamma_{01}})\\
\ket{01} &=& \frac{1}{\sqrt{2}}(\ket{\Gamma_{10}} + \ket{\Gamma_{11}})\\
\ket{10} &=& \frac{1}{\sqrt{2}}(\ket{\Gamma_{10}} - \ket{\Gamma_{11}})\\
\ket{11} &=& \frac{1}{\sqrt{2}}(\ket{\Gamma_{00}} - \ket{\Gamma_{01}}). 
\end{eqnarray*}
The crucial fact to note is that if we fix some $k_1k_2$, then 
there exist exactly two Bell basis vectors $\ket{\Gamma_{i_1i_2}}$ such that 
$|\inp{\Gamma_{i_1i_2}}{k_1k_2}|^2 = 1/2$. For the remaining two basis vectors
the inner product with $\ket{k_1k_2}$ will be zero. 
A simple calculation shows that we can express the two qubit basis states of
the other two mutually unbiased bases analogously: for each two qubit basis state
there are exactly two Bell basis vectors such that the 
inner product is zero and for the other two the inner product squared is $1/2$.

We now take the measurement given by $\{\outp{\Gamma_i}{\Gamma_i}\}$ with 
$\ket{\Gamma_i} = \ket{\Gamma_{i_1i_2}} \otimes \ldots \otimes \ket{\Gamma_{i_{n-1}i_{n}}}$ for the binary
expansion of $i = i_1i_2\ldots i_n$. Fix a $k = k_1k_2\ldots k_n$. By the above argument,
there exist exactly $2^{n/2}$ strings $i \in \01^n$ such that 
$|\inp{\Gamma_i}{k}|^2 = 1/(2^{n/2})$. Putting everything together, Eq.\ (\ref{accessible})
now gives us for any prior distribution $\{p_{t,k}\}$ that
\begin{equation}\label{Ibell}
-\sum_i \bra{\Gamma_i}\mu\ket{\Gamma_i} \log \bra{\Gamma_i}\mu\ket{\Gamma_i} - \frac{n}{2} \leq \mI_{acc}(\ens).
\end{equation}
For our particular distribution we have $\mu = \id/d$ and thus 
$$
\frac{n}{2} \leq \mI_{acc}(\ens).
$$

We now prove a matching upper bound that shows that our measurement is optimal.
For our distribution, we can rewrite Eq.\ (\ref{accessible}) for the POVM 
given by $\{\alpha_i \outp{\Phi_i}{\Phi_i}\}$ to

\begin{eqnarray*}
\mI_{acc}(\ens) &=& \max_M \left(\log d + \right.\\
&&\left. \sum_i \frac{\alpha_i}{d} \sum_{k,t} p_{t} |\bra{\Phi_i}U_t\ket{k}|^2 \log |\bra{\Phi_i}U_t\ket{k}|^2 \right)\\
&=& \max_M \left(\log d -  \sum_i \frac{\alpha_i}{d} \sum_{t} p_t H(\mB_t,\ket{\Phi_i}) \right).
\end{eqnarray*}

It follows from Corollary~\ref{MUderived} that $\forall i\in \{0,1\}^n$ and
$p_1,p_2,p_3\leq 1/2$,
\begin{eqnarray*}
(1/2-p_1) [H(\mB_2,\ket{\Phi_i}) + H(\mB_3,\ket{\Phi_i}) ]&+&\\
(1/2-p_2) [H(\mB_1,\ket{\Phi_i}) + H(\mB_3,\ket{\Phi_i})]&+&\\
(1/2-p_3)[H(\mB_1,\ket{\Phi_i})+H(\mB_2,\ket{\Phi_i})]&& \geq n/2.
\end{eqnarray*}
Reordering the terms we now get
$\sum_{t=1}^3 p_{t} H(\mB_t,\ket{\Phi_i})\geq n/2.$
Putting things together and using the fact that $\sum_i \alpha_i=d$, we obtain
$$
\mI_{acc}(\ens) \leq \frac{n}{2},
$$
from which the result follows.

If, on the other hand, there exists a $t \in [3]$ such that $p_t > 1/2$, then 
by measuring in the basis $\mB_t$ we obtain $\mI_{acc}(\ens) \geq p_t n > n/2$.
\end{proof}

Above, we have only considered a non-uniform prior over the set
of bases. 
In \cite{BalWehWin:pistar} it is observed that when we want to 
guess the XOR of a string of length $2$ encoded in one (unknown to us) 
of these three bases,
the uniform prior on the strings is not the one that gives the smallest
probability of success. This might lead one to think that a similar
phenomenon could be observed in the present setting, i.e., that one
might obtain better locking with three basis for a non-uniform prior on
the strings. In what follows, however, we show that this is not the case.

Let $p_t=\sum_{k} p_{k,t}$ be the marginal distribution on the basis,
then the difference in Bob's knowledge between receiving only the 
quantum state and receiving the quantum state \emph{and} the basis information
is given by 
\begin{eqnarray*}
 \Delta(p_{k,t})=H(p_{k,t})-\mI_{acc}(\ens) -H(p_t),
\end{eqnarray*}
substracting the basis information itself. Consider the post-measurement state
$\nu=\sum_i \bra{\Gamma_i}\mu\ket{\Gamma_i}\ket{\Gamma_i}\bra{\Gamma_i}$. Using (\ref{Ibell}) we obtain
\begin{eqnarray} \label{gap1}
 \Delta(p_{k,t})\leq H(p_{k,t})-S(\nu)+n/2 -H(p_t),
\end{eqnarray}
where $S$ is the von Neuman entropy. Consider the state
\begin{eqnarray*}
\rho_{12} =
\sum_{k=1}^{d} \sum_{t=1}^{3} p_{k,t}(\outp{t}{t})_1
\otimes (U_t \outp{k}{k} U_t^{\dagger})_2,
\end{eqnarray*}
we have that
\begin{eqnarray*}
S(\rho_{12})=H(p_{k,t}) &\leq S(\rho_1) +S(\rho_2)\\
                        & = H(p_t) +S(\mu)\\
                        &\leq H(p_t)+S(\nu).
\end{eqnarray*}
Using (\ref{gap1}) and the previous equation we get
\begin{eqnarray*}
 \Delta(p_{k,t})\leq n/2,
\end{eqnarray*}
for any prior distribution. This bound is saturated by the uniform prior
and therefore we conclude that the uniform prior results in the largest
gap possible. 

\subsection{MUBs from generalized Pauli Matrices}

We first consider MUBs based on the generalized Pauli matrices $X_d$ and $Z_d$ as described
in Section~\ref{prelim}. We consider a uniform prior over the elements of each basis and the set of bases.
Choosing a non-uniform prior does not lead to a better locking effect.

\begin{lemma}\label{equiv}
Let $\mathbb{B}=\{\mB_1,\ldots,\mB_{m}\}$ be any set of MUBs constructed on the basis of generalized Pauli matrices
in a Hilbert space of prime power dimension $d = p^N$.
Consider the ensemble $\ens = \{ \frac{1}{d m},\outp{b^t_k}{b^t_k}\}$. Then
$$
I_{acc}(\ens) = \log d - 
\frac{1}{m} \min_{\ket{\psi}} \sum_{\mB_t \in \mathbb{B}} H(\mB_t,\ket{\psi}).
$$
\end{lemma}
\begin{proof}
We can rewrite Eq.\ (\ref{accessible}) for the POVM
given by $\{\alpha_i \outp{\Phi_i}{\Phi_i}\}$ to
\begin{eqnarray*}
\mI_{acc}(\ens) &=& \max_M \left(\log d + \right.\\
&&\left. \sum_i \frac{\alpha_i}{d m} \sum_{k,t} |\inp{\Phi_i}{b^t_k}|^2 \log |\inp{\Phi_i}{b^t_k}|^2 \right)\\
&=& \max_M \left(\log d -  \sum_i \frac{\alpha_i}{d} \sum_{t} p_{t} H(\mB_t,\ket{\Phi_i}) \right).
\end{eqnarray*}
For convenience, we split up the index $i$ into $i = ab$ with $a = a_1,\ldots,a_N$ and $b=b_1,\ldots,b_N$, 
where $a_\ell,b_\ell \in \{0,\ldots,p-1\}$ in the following.

We first show 
that applying generalized Pauli matrices to the basis vectors of a MUB merely permutes those vectors.
\begin{claim}
Let $\mB_t = \{\ket{b^t_1},\ldots,\ket{b^t_d}\}$ be a basis based on generalized Pauli matrices 
(Section~\ref{prelim}) with
$d = p^N$. Then $\forall a,b \in \{0,\ldots,p-1\}^N, \forall k \in [d]$ we have that $\exists k' \in [d],$ such 
that $\ket{b^{t}_{k'}} = X_d^{a_1}Z_d^{b_1} \otimes \ldots 
\otimes X_d^{a_N}Z_d^{b_N}\ket{b^t_k}$.
\end{claim}
\begin{proof}
Let $\Sigma_p^i$ for $i \in \{0,1,2,3\}$ denote the generalized Pauli's 
$\Sigma_p^0 = \id_p$, 
$\Sigma_p^1 = X_p$, 
$\Sigma_p^3 = Z_p$, and
$\Sigma_p^2 = X_p Z_p$. Note that $X_p^uZ_p^v = \omega^{uv} Z_p^v X_p^u$,
where $\omega = e^{2\pi i/p}$.
Furthermore, define
$
\Sigma_p^{i,(x)} = \id^{\otimes (x - 1)} \otimes \Sigma_p^{i} \otimes \id^{N-x}
$
to be the Pauli operator $\Sigma_p^i$ applied to the $x$-th qupit.
Recall from Section~\ref{prelim} that the basis $\mB_t$ is the unique simultaneous eigenbasis 
of the set of operators in $C_t$, i.e., for all $k \in [d]$ and $f,g \in [N]$,
$\ket{b^t_k} \in \mB_t$ and $c_{f,g}^t \in C_t$, we have
$c_{f,g}^t \ket{b^t_k}=\lambda_{k,f,g}^t \ket{b^t_k} \textrm{ for some value }\lambda^t_{k,f,g}$.
Note that any vector $\ket{v}$ that satisfies this equation
is proportional to a vector in $\mB_t$. To prove
that any application of one of the generalized Paulis merely permutes the vectors in $\mB_t$
is therefore equivalent to proving that $\Sigma^{i,(x)}_{p}
\ket{b^t_k}$ are eigenvectors of $c_{f,g}^t$ for any $f,g \in [k]$ and $i \in \{1,
3\}$. This can be seen as follows: Note that $c_{f,g}^t=\bigotimes_{n=1}^N
\left(\Sigma^{1, (n)}_{p}\right)^{f_N} \left(\Sigma^{3,(n)}_{p}\right)^{g_N}$ 
for $f = (f_1,\ldots,f_N)$ and $g=(g_1, \ldots, g_N)$ 
with $f_N,g_N \in \{0,\ldots,p-1\}$~\cite{boykin:mub}. A calculation then shows that
$$
c_{f,g}^t \Sigma^{i,(x)}_p \ket{b^t_k}= \tau_{f_x,g_x, i} \lambda_{k,f,g}^t \Sigma^{i,(x)}_{p} \ket{b^t_k},
$$
where  $\tau_{f_x,g_x, i}=\omega^{g_x}$ for $i = 1$ and 
$\tau_{f_x,g_x,i}=\omega^{-f_x}$ for $i = 3$. Thus
$\Sigma^{i,(x)}_{p} \ket{b^t_k}$ is an eigenvector of $c^t_{f,g}$ for
all $t, f, g$ and $i$, which proves our claim.
\end{proof}

Suppose we are given $\ket{\psi}$ that minimizes 
$\sum_{\mB_t \in \mathbb{T}} H(\mB_t,\ket{\psi})$. 
We can then construct a full POVM with $d^2$ elements by taking
$\{\frac{1}{d}\outp{\Phi_{ab}}{\Phi_{ab}}\}$ with $\ket{\Phi_{ab}} = (X_d^{a_1}Z_d^{b_1} \otimes \ldots 
\otimes X_d^{a_N}Z_d^{b_N})^\dagger\ket{\psi}$. However, it follows from our claim
above that $\forall a,b,k, \exists k'$ sucht that $|\inp{\Phi_{ab}}{b^t_k}|^2 = |\inp{\psi}{b^{t}_{k'}}|^2$, 
and thus
$H(\mB_t,\ket{\psi}) = H(\mB,\ket{\Phi_{ab}})$ from which the result follows.
\end{proof}

Determining the strength of the locking effects for such MUBs is thus equivalent to proving bounds on 
entropic uncertainty relations. We thus obtain as a corollary of 
Theorem~\ref{squareThm} and Lemma~\ref{equiv}, that, for dimensions which are the square of a prime power 
$d = p^{2N}$, using any product MUBs based on generalized Paulis does not give us any better
locking than just using 2 MUBs. 
\begin{corollary}\label{pauliLocking}
Let $\mathbb{S}=\{\mS_1,\ldots,\mS_{m}\}$ with $m \geq 2$ be any set of MUBs constructed on the basis of generalized Pauli matrices
in a Hilbert space of prime (power) dimension $s = p^N$. 
Define $U_t$ as the unitary that transforms the computational basis
into the $t$-th MUB, i.e., $\mS_t = \{U_t\ket{1},\ldots,U_t\ket{s}\}$. 
Let $\mathbb{B} = \{\mB_1,\ldots,\mB_{m}\}$ be the set of product MUBs with
$\mB_t = \{U_t \otimes U_t^* \ket{1},\ldots,U_t \otimes U_t^*\ket{d}\}$ in dimension $d=s^2$.
Consider the ensemble $\ens = \{ \frac{1}{d m},\outp{b^t_k}{b^t_k}\}$. Then
$$
I_{acc}(\ens) = \frac{\log d}{2}.
$$
\end{corollary}
\begin{proof}
The claim follows from Theorem~\ref{squareThm} and the proof of Lemma~\ref{equiv}, by constructing
a similar measurement formed from vectors $\ket{\hat{\Phi}_{\hat{a}\hat{b}}} = K_{a^1b^1} 
\otimes K_{a^2b^2}^* \ket{\psi}$
with $\hat{a} = a^1a^2$ and $\hat{b} = b^1b^2$, where $a^1,a^2$ and $b^1,b^2$ are
defined like $a$ and $b$ in the proof of Lemma~\ref{equiv}, and $K_{ab} = (X_d^{a_1}Z_d^{b_1}\otimes\ldots\otimes X_d^{a_N}Z^{b_N}_d)^\dagger$
from above.
\end{proof}

The simple example we considered above is in fact a special case of Corollary~\ref{pauliLocking}. 
It shows that if the vector that minimizes the sum of entropies has certain symmetries, 
such as for example the Bell states, the resulting POVM can even be much simpler.

\subsection{MUBs from Latin Squares}

At first glance, one might think that maybe the product MUBs based on generalized Paulis are not well suited
for locking just because of their product form. Perhaps MUBs with entangled basis vectors do not exhibit this problem.
To this end, we examine how well MUBs based on Latin squares can lock classical information in a quantum state.
All such MUBs are highly entangled, with the exception of the two extra MUBs based on non-Latin squares.
Surprisingly, it turns out, however, that \emph{any} set of at least two MUBs based on Latin squares, does equally well
at locking as using just 2 such MUBs. Thus such MUBs perform equally ``badly'', i.e., we cannot improve the strength of 
the locking effect by using more MUBs of this type. 
   
\begin{lemma}\label{LSlocking}
Let $\mathbb{B}=\{\mB_1,\ldots,\mB_m\}$ with $m \geq 2$ be any set of MUBs in a Hilbert space of dimension $d=s^2$ constructed
on the basis of Latin squares.
Consider the ensemble $\ens = \{ \frac{1}{d m},\outp{b^t_k}{b^t_k}\}$. Then
$$
\mI_{acc}(\ens) = \frac{\log d}{2}.
$$ 
\end{lemma}
\begin{proof}
Note that we can again rewrite $\mI_{acc}(\ens)$ as in the proof of Lemma~\ref{equiv}. Consider 
the simple measurement in the computational basis $\{\outp{i,j}{i,j}|i,j \in [s]\}$. The result
then follows by the same
argument as in Lemma~\ref{LSentropy}.
\end{proof}

\section{Conclusion and Open Questions}

We have shown tight bounds on entropic uncertainty relations and locking for specific sets of mutually unbiased bases. 
Surprisingly, it turns out that using more mutually unbiased basis does not always lead to a better locking effect. 
It is interesting to consider what may make these bases so special. The example of three MUBs considered in Lemma~\ref{3mubLocking}
may provide a clue. These three bases are given by the common eigenbases of $\{\sigma_x \otimes \sigma_x, \sigma_x \otimes \id,
\id \otimes \sigma_x\}$, $\{\sigma_z \otimes \sigma_z, \sigma_z \otimes \id, \id \otimes \sigma_z\}$ and $\{\sigma_y \otimes \sigma_y,
\sigma_y \otimes \id, \id \otimes \sigma_y\}$ respectively~\cite{boykin:mub}. However, $\sigma_x \otimes \sigma_x$, $\sigma_z \otimes \sigma_z$ and
$\sigma_y \otimes \sigma_y$ commute and thus also share a common eigenbasis, namely the Bell basis. This is exactly the basis we will use as our
measurement. For all MUBs based on generalized Pauli matrices, the MUBs in prime power dimensions are given as the common eigenbasis
of similar sets consisting of strings of Paulis. It would be interesting to determine the strength of the locking effect on the
basis of the commutation relations of elements of \emph{different} sets. Perhaps it is possible to obtain good locking from a subset of such MUBs
where none of the elements from different sets commute.

It is also worth noting that  the numerics of~\cite{terhal:locking}
indicate that at least in dimension $p$ using more than three bases does indeed lead to a stronger
locking effect. It would be interesting to know, whether the strength of the locking effect
depends not only on the number of bases, but also on the dimension of the system in question.

Whereas general bounds still elude us, we have shown that merely choosing mutually unbiased bases is not sufficient to obtain good locking effects or high lower bounds
for entropic uncertainty relations.  We thus have to look for different properties.  
\acknowledgments
We would like to thank Harry Buhrman, Hartwig Bosse, Matthias Christandl, Richard Cleve, Debbie Leung, Serge Massar, 
David Poulin, and Ben Toner for discussions. We would especially like to thank Andris Ambainis and Andreas Winter for many 
helpful comments and interesting discussions. We would also like to thank Debbie Leung, John Smolin and Barbara Terhal for providing 
us with explicit details on the numerical studies conducted in~\cite{terhal:locking}.
Thanks also to Matthias Christandl and Serge Massar for discussions on errors in string commitment protocols, to which
end claim 1 was proved in the first place. Thanks also to Matthias Christandl and Ronald de Wolf for helpful 
comments on an earlier version of this note. 
We are supported by an NWO vici grant 2004-2009
and by the EU project QAP (IST 015848).

\end{document}